\documentclass[aps,prl,twocolumn,floatfix]{revtex4}
\usepackage{amsmath}
\usepackage{graphicx}		
\usepackage{graphics}
\usepackage[usenames]{color}

\begin{document}

\newcommand{\diff}[2]{\frac{\partial #1}{\partial #2}}
\newcommand{\secdiff}[2]{\frac{\partial^2 #1}{\partial #2^2}}
\newcommand{\comment}[1]{{\bf \textcolor{red}{ #1}}}
\newcommand{\NPMcomment}[1]{{\bf \textcolor{ForestGreen}{ #1}}}
\newcommand{\twovec}[2]{\left(\begin{array}{c} 
#1 \\ #2 \end{array}\right)}
\newcommand{\threevec}[3]{\left(\begin{array}{c} 
#1 \\ #2 \\ #3 \end{array}\right)}
\newcommand{\twomatrix}[4]{\left(\begin{array}{cc} 
#1 & #2 \\ #3 & #4 \end{array}\right)}
\newcommand{\threematrix}[9]{\left(\begin{array}{ccc} 
#1 & #2 & #3 \\ #4 & #5 & #6 \\ #7 & #8 & #9 \end{array}\right)}
\newcommand{\twomatrixdet}[4]{\left|\begin{array}{cc} 
#1 & #2 \\ #3 & #4 \end{array}\right|}

\title{Efimov states embedded in the three-body continuum}

\author{N.~P.~Mehta}
\email[]{mehtan@jilau1.colorado.edu}
\affiliation{Department of Physics and JILA, University of Colorado, Boulder, CO 80309-0440}
\author{Seth~T.~Rittenhouse}
\email[]{rittenho@colorado.edu}
\affiliation{Department of Physics and JILA, University of Colorado, Boulder, CO 80309-0440}
\author{J.~P.~D'Incao}
\email[]{jpdincao@jilau1.colorado.edu}
\affiliation{Department of Physics and JILA, University of Colorado, Boulder, CO 80309-0440}
\author{Chris~H.~Greene}
\email[]{chris.greene@colorado.edu}
\affiliation{Department of Physics and JILA, University of Colorado, Boulder, CO 80309-0440}

\date{\today}

\begin{abstract}
We consider a multichannel generalization of the Fermi pseudopotential to model 
low-energy atom-atom interactions near a magnetically tunable Feshbach resonance, 
and calculate the adiabatic hyperspherical potential curves for a system of 
three such interacting atoms.  In particular, our model suggests the
existence of a series of quasi-bound Efimov states attached to excited
three-body thresholds, far above open channel collision energies.  We
discuss the conditions under which such states may be supported, and
identify which interaction parameters limit the lifetime of these states.
We speculate that it may be possible to observe these states using
spectroscopic methods, perhaps allowing for the measurement of multiple
Efimov resonances for the first time.
\end{abstract}


\maketitle


Three resonantly interacting particles may form long-range bound states, called 
Efimov states~\cite{Efimov1}, even if the short-range interparticle interaction 
supports no two-body bound states.  
After years of tantalizing work on atomic ${}^4$He trimers 
and halo nuclei (see~\cite{braaten2006ufb} and references therein),
the first strong experimental evidence of Efimov physics was recently found in an ultracold 
thermal gas of cesium atoms~\cite{KraemerNature2006}.  This observation was
made possible by utilizing a magnetically tunable Feshbach resonance to 
precisely control the two-body scattering length over a substantial range,
and observing a resonant feature in the atom-loss rate due to three-body
recombination~\cite{EsryGreeneBurke1999,NielsenMacek,BraatenHammer2001PRL}.
Since hyperspherical studies have provided important insights in the 
past~\cite{NielsenEtAlPhysRep2001,EsryGreeneBurke1999,DincaoEsry2005}, we
are motivated in this Letter to investigate the nature of universality within the 
adiabatic hyperspherical representation for multichannel two-body interactions, 
obtaining a complete view of the complicated energy landscape. 

Multichannel systems have a considerably richer structure than their single-channel 
counterparts~\cite{BulgacEfimov}, admitting multiple three-body thresholds, 
quasi-bound two-body channels, and multiple length scales.  In this Letter, 
we present a Green's function method for obtaining the adiabatic 
potential curves, and discuss some immediate consequences.  
Note that the zero-range multichannel method developed initially by Macek
and Kartavtsev to study three-body recombination near a narrow two-body
Feshbach resonance~\cite{KartavtsevMacekFBS2002} is the closest existing study 
to the present analysis, but it did not consider our main 
point of focus, namely the Efimov effect for \emph{excited} three-body thresholds.

\begin{figure}
\leavevmode
\includegraphics[width=2.2in,clip=true]{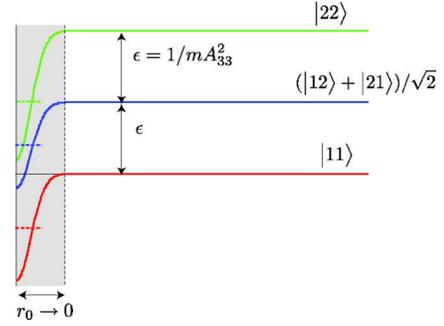}
\caption{(color online) A schematic representation of our two-body model is
shown.  As the range $r_0$ is taken to zero, the potential is regularized
by the derivative in Eq.~(\ref{v2b})}.
\label{fig:2bodymultichannel1}
\end{figure}

Efimov states attached to \emph{excited} three-body thresholds 
are absent in single channel models and difficult to infer from
other multichannel approaches~\cite{LeeKohlerJulienne}.  For these 
excited thresholds, at least one of the three atoms resides in an 
excited Zeeman or hyperfine state.  We label these one-atom states 
$|1\rangle$ and $|2\rangle$ with energies $E_1=0$ and $E_2=\epsilon$, 
respectively.  In particular, we show that under the rather general 
stipulation that a quasibound two-body state is degenerate with an excited
two-body threshold (see Fig.~\ref{fig:2bodymultichannel1}), 
and that the coupling between product states $|11\rangle$ 
and $|22\rangle$ is much smaller than all other couplings, a series of 
Efimov states should emerge that is attached to an excited three-body threshold. 
These Efimov states are of a unique character and can be viewed as fully three-body 
Fano-Feshbach resonances embedded within the three-body continuum far above 
the energy of open channel collisions, in contrast with the single-channel 
Efimov resonances predicted and recently observed that were interpreted as shape 
resonances~\cite{EsryGreeneBurke1999,NielsenMacek}.  Since the interatomic scattering length for ground-state atoms will not
in general be resonant, the gas will be comparatively stable with respect to the 
$a^4$ scaling law for three-body recombination~\cite{EsryGreeneBurke1999,NielsenMacek,braaten2006ufb}.  
Efimov states attached to the excited threshold could be directly 
accessed via photoassociation and observed through the measurement of atom-loss 
rates, thus opening a new toolkit of powerful spectroscopic 
techniques~\cite{PAReferences}.

As illustrated in Fig.~\ref{fig:2bodymultichannel1}, the two-body system
has a set of product states  
$\{|\sigma=1\rangle,|\sigma=2\rangle,|\sigma=3\rangle \} = \{|11\rangle, 
(|12\rangle + |21\rangle)/\sqrt{2}, |22\rangle\}$
with respective energies $\{E_\sigma\}=\{0,\epsilon,2\epsilon\}$.  
The zero-range two-body potential is now written in the $\{|\sigma\rangle\}$ basis as,
\begin{equation}
\label{v2b}
\underline{V}(r)=\frac{4 \pi \underline{A}}{m}\frac{\delta(r)}{r^2}\diff{}{r}(r 
\; \cdot),
\end{equation}
This potential is the natural extension of the regularized Fermi pseudopotential 
which imposes the Bethe-Peierls boundary condition on the two-body wavefunction, 
$\psi(r) \rightarrow C(1-a/r)$ as $r\rightarrow 0$.  When all channels are 
energetically open, the matrix $\underline{A}$ is simply minus the scaled 
two-body reaction matrix $\underline{K}_0$.  The physical reaction matrix
is obtained from the multichannel quantum defect theory (MQDT) channel
closing formula~\cite{RossShawAnnPhys1961,BurkePRL1998}.  Note that unlike 
other multichannel treatments that retain only one open background channel 
and a one resonant molecule channel~\cite{LeeKohlerJulienne}, we label our 
channels explicitly by the internal states of the atoms, and our
Hamiltonian allows for scattering at excited thresholds between atoms with 
different internal states.


Unless explicitly stated, we choose the energy independent symmetric matrix
to have elements $A_{ij}$ such that $A_{13}\rightarrow 0$, and $A_{33}=A_{11}\sim3.75~a_0$ giving 
$\epsilon/h = \hbar^2/(h m A_{33}^2)\sim 3$~GHz and placing a quasi-bound state at 
$E=\epsilon$.  The quantity $A_{33}$ is the length scale by which we measure all other 
quantities for the remainder of this Letter.
We choose $A_{12}=A_{23}=A_{33}/4$ to give a reasonable resonance width 
$\Delta_B\sim~-8$G, and $A_{22}=2A_{33}$ in order to illustrate the effect
of a quasi-bound two-body state on the three-body potential energy curves.
These numbers are chosen to loosely model the $155$G resonance in ${}^{85}$Rb.  
Explicitly allowing $A_{23}\ne A_{12}$ gives qualitatively similar results in 
the three-body calculations described below.  Allowing $A_{13} \ne 0$ provides a direct 
coupling between $|11\rangle$ and $|22\rangle$ states
and, as we discuss later, this coupling element is expected to limit the 
lifetime of Efimov states near $E=\epsilon$.  We note that the 
methods of~\cite{BurkePRL1998} permit the application of these ideas over a wider 
energy range because they include the long range van der Waals physics in the 
QDT treatment explicitly.  

In the three-body system, we transform the mass-scaled Jacobi coordinates: 
$\vec{y}^{(i)}_1 = (\vec{r}_j - \vec{r}_k)/d$
and $\vec{y}^{(i)}_2=d(\frac{\vec{r}_j+\vec{r}_k}{2}-\vec{r}_i)$ (with 
$d=2^{1/2}/3^{1/4}$) into hyperspherical coordinates by defining the
hyperangle $\alpha$ as $\tan{\alpha^{(i)}}=|y_1^{(j)}|/|y_1^{(k)}|$, the hyperradius $R$ as 
$R^2=\vec{y}_1^2 + \vec{y}_2^2$ (which is invariant
under permutations) and the spherical polar angles 
${\omega_i}=\{\theta_i,\phi_i\}$ point in the direction of $\hat{y}_i$.  
The angular coordinates $\{\alpha, \omega_1, \omega_2\}$ are collectively 
denoted $\Omega$.
The pair-wise interactions are written in the basis of two-body internal states, 
with the third spectator
particle (near any chosen two-particle coalescence point) in either $|1\rangle$ 
or $|2\rangle$.  Hence, if particle $1$ is the spectator particle, the basis of 
internal states is $\{|\Sigma\rangle\} = \{|111\rangle, (|112\rangle + |121\rangle)/\sqrt{2}, 
|211\rangle, |122\rangle, (|212\rangle + |221\rangle)/\sqrt{2},|222\rangle \}$ with energies 
$\{E_\Sigma\}=\{0, \epsilon,\epsilon,2\epsilon,2\epsilon, 3\epsilon \}$.

In the multichannel generalization~\cite{KartavtsevMacekFBS2002} of the adiabatic 
hyperspherical method~\cite{MacekJPB1968},  we seek solutions to:
\begin{equation}
\left(\frac{\Lambda^2}{2 \mu R^2} + \underline{E_{th}} + 
\underline{V}(R,\Omega)\right)\vec{\Phi}(R;\Omega) = 
U(R) \vec{\Phi}(R;\Omega)
\end{equation}
Here, $\underline{E}_{th}$ is a diagonal matrix 
$[E_{th}]_{\Sigma\Sigma'}=\delta_{\Sigma\Sigma'}E_\Sigma$, and
$\underline{V}(R,\Omega)$ is the sum of matrices for each 
pair-wise interaction expressed in the basis $\{|\Sigma\rangle\}$.  The
three-body reduced mass is $\mu=m/\sqrt{3}$.
The adiabatic potential is written in terms of the eigenvalue $\nu=\nu_1$ as 
$U(R)=\nu_1(\nu_1+4)/2\mu R^2$, where
$\nu_\Sigma(\nu_\Sigma+4) = 2 \mu R^2(U(R)-E_\Sigma)$.
Components of the free-space (diagonal) hyperangular Green's 
function $\underline{G}(\Omega,\Omega')$ satisfy,
\begin{equation}
\label{GFeq}
\left(\Lambda^2-\nu_{\Sigma}(\nu_{\Sigma}+4)\right)G_{\Sigma 
\Sigma}(\Omega,\Omega') = \delta(\Omega,\Omega')
\end{equation}
Defining $g^{\nu_\Sigma}_{l_1 l_2}(\alpha,\alpha')=N_{\nu_\Sigma l_1 l_2}
f^{(-)}_{\nu_\Sigma l_1 l_2}(\alpha_<)f^{(+)}_{\nu_\Sigma l_1 l_2}(\alpha_>)$,
the most useful solution is written in the Sturm-Liouville form~\cite{MehtaRittenhouseUnpublished},
\begin{align}
G_{\Sigma\Sigma}(\Omega,\Omega')&=\sum_{l_1,m_1,l_2,m_2}[\;g^{\nu_\Sigma}_{l_1
l_2}(\alpha,\alpha') \times \notag \\ 
&Y_{l_1 m_1}(\omega_1)Y^*_{l_1 m_1}(\omega_1')Y_{l_2 m_2}(\omega_2)Y^*_{l_2 
m_2}(\omega_2')\big ]
\end{align}
where $f^-$ ($f^+$) is a solution to the homogeneous version of
Eq.~(\ref{GFeq}) regular at $\alpha=0$ ($\alpha=\pi/2$), and
the normalization $N_{\nu_\Sigma l_1 l_2}$ is fixed by the Wronskian of
$f^-$ and $f^+$~\cite{JacksonEM}.
The hyperangular Lippmann-Schwinger (L-S) equation,
\begin{equation}
\Phi_\Sigma(\Omega)=-2\mu R^2 \sum_{\Sigma',k}\int d\Omega'\; G_{\Sigma 
\Sigma}(\Omega,\Omega')V^{(k)}_{\Sigma \Sigma'}(R,\Omega')
\Phi_{\Sigma'}(\Omega')
\end{equation}
is then solved by evaluating the integral over $V^{(k)}_{\Sigma\Sigma'}$ in the coordinate system 
where $|\vec{y}^{(k)}_1| \propto |\vec{r}_i -\vec{r}_j| \propto 
R\sin{\alpha^{(k)}}$.  The zero-range s-wave interaction immediately gives $l_1=0$, and $l_2=L$.  For this Letter, 
we confine our study to states with total orbital angular momentum $L=0$.  For 
equal-mass particles, in the limit $\alpha^{(k)} \rightarrow 0$ we have 
$\alpha^{(i)}=\alpha^{(j)}=\pi/3$; the L-S equation reduces to a matrix
equation of the form~\cite{KartavtsevMacekFBS2002}:
\begin{equation}
\label{det}
\left(\frac{3^{1/4}}{2^{1/2} R}\left(\underline{M}^{(1)} + \underline{M}^{(2)} 
\underline{P_-} + 
\underline{M}^{(3)} \underline{P_+}\right) -
\underline{1}\right)\vec{C}^{(1)}=0
\end{equation}
where, for bosons,
\begin{equation}
M^{(i)}_{\Sigma \Sigma'} = 
\begin{cases}
A^{(i)}_{\Sigma \Sigma'} \lambda_\Sigma 
\cot{(\lambda_\Sigma \pi/2)}  & ~~ i=1 \\
A^{(i)}_{\Sigma \Sigma'} \frac{-4 
\sin{(\lambda_\Sigma \pi/6)}}{\sqrt{3} \sin{(\lambda_\Sigma \pi/2)}} & ~~ i=2,3
\end{cases}
\end{equation}
The vector associated with the $n^{\text{th}}$ eigenstate, $\vec{C}^{(1)}_n$
in Eq.~(\ref{det}), is defined by the boundary 
and normalization conditions:
\begin{equation}
\label{Cvec2}
C^{(k)}_{\Sigma,n}=\lim_{r_k \rightarrow 0}{\diff{(r_k\Phi_{\Sigma,n})}{r_k}} ~~~ 
{\rm and} ~~~ \sum_\Sigma{\langle \Phi_{\Sigma,n}|\Phi_{\Sigma,n}\rangle}=1.
\end{equation}
We have rewritten the eigenvalue as $\nu_\Sigma=\lambda_\Sigma-2$ purely for 
convenience, and $\underline{P_+}$ and $\underline{P_-}$ perform cyclic and anticyclic permutations respectively upon the 
basis $\{|\Sigma\rangle\}$. Zeros of the determinant of the
matrix in Eq.~(\ref{det}) yield the eigenvalues $\lambda_{\Sigma,n}^2=2\mu 
R^2 (U_n(R)-E_\Sigma)+1/4$.  
The potentials $U_n(R)$ appear in radial equations of the form (ignoring
nonadiabatic effects): $-F_n''(R)/2\mu R^2 + U_n(R)F_n(R)=E F_n(R)$.

\begin{figure}
\leavevmode
\includegraphics[width=2.5 in,clip=true]{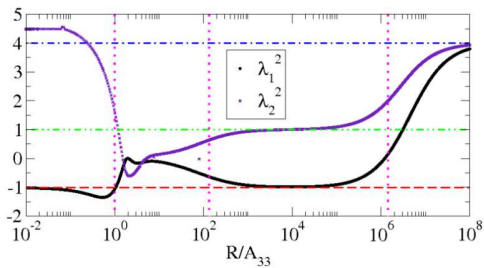}
\caption{(color online) The adiabatic eigenvalues $\lambda_1^2$ ($\lambda_2^2$) with
respect to the first (second) threshold are shown for a near-resonant
choice of $\epsilon$.  From left to right, the vertical dotted magenta lines mark $A_{33}$, 
$r_e=132.284 A_{33}$, and $a_{open}=-1.445\times 10^6 A_{33}$.  The horizontal 
dashed red line corresponds to the
universal eigenvalue for three equal-mass resonantly interacting 
bosons $\lambda_1^2=-1.012514$~\cite{Efimov1}, while the green dash-dot-dot line 
indicates the universal eigenvalue
for only one resonant interaction $\lambda^2=1$, and the horizontal dash-dot 
blue line marks the lowest large $R$ free-space eigenvalue $\lambda^2=4$.}
\label{fig:LenScales2}
\end{figure}

In Fig.~\ref{fig:LenScales2}, we show eigenvalues with respect to the $E=0$
and $E=\epsilon$ thresholds for $\epsilon$ chosen so that a quasi-bound two-body 
state is nearly degenerate with the $E=0$ threshold, giving an 
identical boson (open-channel) scattering length of 
$a_{open} = -1.445 \times 10^6 A_{33}$.  Note that we now drop the index $n$
from the label of $\lambda$ since the curves shown in
Fig.~\ref{fig:LenScales2} are actually comprised of many distinct
eigenvalues connected by a series of close avoided crossings.  This figure illustrates 
how the multichannel problem naturally introduces different length 
scales and symmetry constraints to the three-body problem.  The 
eigenvalues may be understood by considering simultaneously the internal
states of the atoms and the important natural length scales: $A_{33}$, the 
open-channel effective range $r_e$, and the open-channel scattering 
length $a_{open}$.  When $R \ll A_{33}$, all length scales in our 
model appear long, so with respect to the lowest threshold, we have 
$\lambda_1^2 \rightarrow -1.012515$ appropriate for three resonantly 
interacting bosons.  The region $A_{33} \ll R \ll r_e$ is a 
transitional regime indicating the importance of the $r_e$ length-scale.
In the region $r_e \ll R \ll \left| a_{open} \right|$, the potential near $E=0$ behaves again 
as that of three resonantly interacting bosons, while near the second 
threshold $E=\epsilon$, only one of the interactions appears resonant 
and we obtain the appropriate value $\lambda_2^2=1$~\cite{DincaoEsry2007}.  For 
$\left| a_{open} \right| \ll R$, the eigenvalues approach the free-space value 
$\lambda_1^2=\lambda_2^2=4$.  Note that for values of $R$ where $\lambda_1^2 < 0$
($\lambda_2^2 < 0$) and is approximately constant, the potential with
respect to the $E=0$ ($E=\epsilon$) threshold is supercritical and 
supports Efimov states.  

\begin{figure*}
\leavevmode
\includegraphics[width=6.0 in, clip=true]{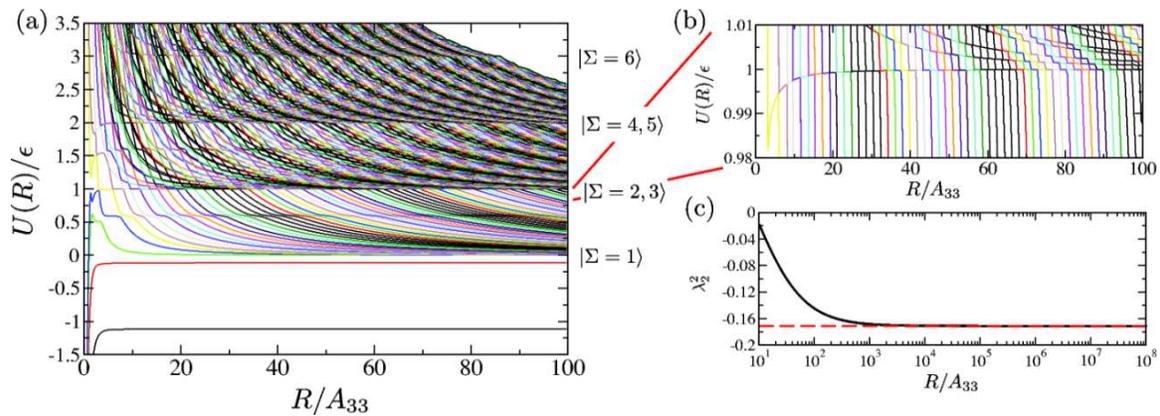}
\caption{
In (a), we show approximately the lowest $300$ potential curves up to 
$3.5 \epsilon$, while (b) shows an
enlarged view of the region  near the $E=\epsilon$ threshold, and the
attractive $R^{-2}$ diabatic Efimov potential.  Note also the series of avoided
crossings in (a) near $E=0.6 \epsilon$ indicating the presence of a two-body
quasi-bound state.  In (c) we show the eigenvalue near $E=\epsilon$
converging to the universal value for two identical bosons and one
distinguishable particle $\lambda^2 \rightarrow -0.171145$ indicated by the
dashed red line.}
\label{fig:AllCurves3}
\end{figure*}

As mentioned before, by considering $\epsilon$ to vary with the magnetic field, it is possible to 
make a two-body state degenerate with the $|\sigma = 2\rangle$ threshold.
This occurs when $\epsilon \rightarrow 1/(m A_{33}^2)$ as illustrated in Fig.~\ref{fig:2bodymultichannel1}.
Provided that $\left| A_{13} \right| \ll A_{33}$, an Efimov
potential appears at \emph{both} the $E=\epsilon$ and $E=2\epsilon$ thresholds under these conditions.
In Fig.~\ref{fig:AllCurves3}(a) we show the lowest 300 adiabatic potential curves up to
$3.5 \epsilon$.  Note the collection of three-body curves converging to each
three-body threshold at $E=E_\Sigma$. 
Note also the quasi-bound two-body threshold near $0.6 \epsilon$.  A
three-body collision near this energy will couple strongly to the two-body
Feshbach resonance through this 
series of avoided crossings. 
In Fig.~\ref{fig:AllCurves3}(c), the eigenvalue $\lambda_2^2=\lambda_1^2-2\mu R^2 \epsilon$ is shown to
converge to the universal value
$\lambda_2^2=-0.171145$ predicted by the purely imaginary root of 
the corresponding single channel equation in the limit 
$\left| a \right|\rightarrow  \infty$~\cite{Efimov1973}
appropriate for two identical bosons interacting resonantly with
a third distinguishable particle of equal mass, consistent with the fact 
that the $E=\epsilon$ threshold corresponds to internal states of 
the form 
$|\Sigma=2\rangle = (|112\rangle - |121\rangle)/\sqrt{2}$ and $|\Sigma=3\rangle= 
|211\rangle$. 
The spacing of Efimov states obeys the formula $E_n=E_0 \exp{(-2 \pi n/s_0)}$, 
where $\lambda^2 \rightarrow -s_0^2$ in the universal regime, and $E_0$ is 
the energy of the lowest Efimov state.  If we simply consider the 
value of the Efimov potential at $R\sim 100 A_{33}$ to be a 
reasonable estimate for the energy of the first Efimov state, we 
obtain $E_0/h \sim (\epsilon/h)-(10 \text{kHz})$.  This places the second state
at approximately $E_2/h\sim (\epsilon/h)-(1 \text{mHz})$, giving a separation of roughly 
$10$~kHz, within the resolution of modern radio-frequency 
experiments~\cite{PAReferences}.
The spacing between states could be made more favorable by using heteronuclear 
mixtures of atoms, capitalizing on mass-ratios different from 
unity.~\cite{Efimov1,Efimov1973,DincaoEsry2006}.

In order for the Efimov states to persist at long-range, it is also necessary to 
show that the positive-definite non-adiabatic diagonal coupling 
$-Q(R)=-\langle \Phi(R)|\secdiff{}{R}|\Phi(R) \rangle$ corresponding
to the Efimov potential falls off faster than $R^{-2}$. 
This is a non-trivial task since the potential 
is comprised of all potential curves approaching the lowest three-body threshold.  
We derive an expression for $-Q(R)$~\cite{MehtaRittenhouseUnpublished} via a  
multichannel generalization of the method used
in~\cite{KartavtsevMalykh2006}.  Our calculations
indicate that the vector $\vec{C}^{(1)}$
(normalized according to Eq.~(\ref{Cvec2})) for the eigenvalue 
corresponding to the Efimov potential has a dominant component:
$C^{(1)}_{\Sigma=4}$, which grows linearly with $R$.  
Extrapolating this behavior for $\vec{C}^{(1)}$ and using the fact that 
$\lambda_2^2 \rightarrow -0.171145$ as $R^{-1}$,
the expression for $-Q(R)$ reduces to the simple form:
$-Q(R) =3^{3/4} \sqrt{2}\;\pi^2\frac{A_{33}^3 K_4^2}{R^3}$,
where $K_4$ is the slope of the linearly increasing component $C^{(1)}_4$.  Since
$-Q(R)$ falls off faster than $R^{-2}$, we expect these states to persist
at long range.

The lifetime of Efimov trimers near $E=\epsilon$ is limited by 
inelastic processes at short and long range.  Using a WKB estimate, if
all inelastic transitions occur at short range with probability $P_{SR}$ 
(a reasonable assumption in the limit $A_{13}\rightarrow 0$), then 
the width of successive resonances will scale geometrically with the
$n^\text{th}$ energy level as $\Gamma_{SR}\approx P_{SR}E_0 e^{-2n\pi/s_0}/s_0$.

For finite $A_{13}$ inelastic transitions can also occur at long range due to two-body inelastic transitions, and
the width of Efimov resonances is dominated by the size of 
$A_{13}$.  Letting $A_{13} \ne 0$ serves to
broaden the avoided crossing comprising the universal Efimov potential in
Fig.~\ref{fig:AllCurves3}(b), dramatically increasing the decay rate to
three-body continuum states associated with the $|\Sigma=1\rangle$ state,
and making the Efimov states shorter lived and more difficult to observe.
This is understood by noting that when the coupling between $|\sigma=3\rangle$ and $|\sigma=1\rangle$ is significant
(i.e. $A_{13}$ is sizable), direct inelastic transitions between $|\sigma=3\rangle$ 
and $|\sigma=1\rangle$ states dominate over 
inelastic transitions via the intermediate $|\sigma=2\rangle$ state.  

To summarize, we have identified a new class of Efimov states embedded in
the three-body continuum.  The hyperradial potentials supporting these
states have universal properties consistent with the symmetry constraints
implied by the internal degrees of freedom of the three-atom system, and
can further be understood in terms of relevant two-body length scales.    
Further, we have identified the coupling constant 
$A_{13}$ responsible for limiting the lifetime of these Efimov states at
long range, and have shown that the repulsive diagonal nonadiabatic 
correction falls off as $R^{-3}$, consistent with the single-channel
result.  We stress that since the potential supporting these states appears
when a quasi-bound two-body state is 
degenerate with an \emph{excited} two-body threshold, the
open-channel scattering length in general will not be resonant, and the gas 
is expected to be stable with respect to the $a^4$ recombination
scaling law.  We postulate that it may be possible to observe these states by 
spectroscopic techniques, perhaps with sufficient accuracy to
measure two Efimov resonances for the first time.  

We thank D.~Blume and J.~H.~Macek for useful discussions
during the early stages of this work.  This work is supported by the
National Science Foundation.

\end{document}